\documentclass[prl,twocolumn,showpacs,preprintnumbers,amsmath,amssymb,superscriptaddress,floatfix,showkeys]{revtex4}

\usepackage{graphicx}
\usepackage{dcolumn}
\usepackage{bm}

\usepackage{amsfonts}
\usepackage{amsmath}
\usepackage{amsthm}

\begin{document} 
 
 
 \title{The Effects of Inter-particle Attractions on Colloidal Sedimentation}
 
\author{A.\ Moncho-Jord\'{a}} 
\affiliation{Departamento de F\'{\i}sica Aplicada, Facultad de Ciencias, Universidad de Granada, Campus Fuentenueva S/N, 18071 Granada, Spain.}

\author{A. A. Louis} 
\affiliation{Rudolf Peierls Centre for Theoretical Physics, 1 Keble Road, Oxford OX1 3NP, United Kingdom}

\author{J. T. Padding} 
\affiliation{Computational Biophysics, University of Twente, P.O. Box 217, 7500 AE, Enschede, The Netherlands}

 
\begin{abstract}
We use a mesoscopic simulation technique to study the effect of short-ranged inter-particle attraction on the steady-state sedimentation of colloidal suspensions. Attractions increase the average sedimentation velocity $v_s$ compared to the pure hard-sphere case, and for strong enough attractions, a non-monotonic dependence on the packing fraction $\phi$ with a maximum velocity at intermediate $\phi$ is observed.
Attractions also strongly enhance hydrodynamic velocity fluctuations, which show a pronounced maximum size as a function of $\phi$. These results are linked to a  complex interplay between hydrodynamics and the formation and break-up of transient many-particle clusters.
\end{abstract} 

\maketitle 
 
The steady state sedimentation of a set of spherical particles through a viscous solvent at low Reynolds number is a classic problem in non-equilibrium physics that, despite its apparent simplicity, it is still far from completely understood ~\cite{Rama01,Russ89}.  A major difficulty for theories arises from the long-ranged ($1/r$) nature of the solvent induced hydrodynamic interactions (HI) that couple the motion of the particles in a complex way. Indeed, although Stokes ~\cite{Stok51} calculated the sedimentation velocity $v_s^0$ of a single sphere in 1851, it was necessary to wait more than 120 years until Batchelor ~\cite{Batc72} showed that the first correction for a hard sphere system in the dilute limit is given by $v_s/v_s^0=1-6.55\phi$, where $\phi$ is the particle volume fraction.
Although the sedimentation of one or two particles can be solved analytically ~\cite{Russ89}, the hydrodynamic problem of three or more sedimenting particles is intrinsically chaotic ~\cite{Jano97}.    Moreover, simple scaling arguments suggest that for a larger number of particles,  the hydrodynamic velocity fluctuations  around the average $\delta v = v - v_s$  should diverge with container size $L$ as $\left< (\delta v)^2\right> \sim L$~\cite{Cafl85}.
 These fluctuations have a purely hydrodynamic origin and are induced by the chaotic trajectory of the particles, leading to velocity correlations (swirls) that propagate over large distances.  Experiments show that the swirls   grow with container size for smaller $L$ (the unscreened regime) and then saturate for larger containers (the screened regime)~\cite{Segr97}, but the nature and origins of the screening are still a source of controversy~\cite{Rama01,Koch91,Ladd02,Much04}.
  
 Colloidal sedimentation has widespread relevance in industrial applications such as paints, coatings, ceramics, food, and cosmetics. Inter-particle interactions are critical to tuning the desired dispersion properties.  Nevertheless, 
  most theoretical work has focussed on purely repulsive hard sphere (HS) particles, and much less is known about the effect of attractions on sedimentation.
Batchelor's extended his calculations~\cite{Batc72} to include interactions beyond the HS limit~\cite{Batc82a}; for short-ranged  potentials  this can be approximated as~\cite{Russ89}:
\begin{equation}\label{eq:eq1}
v_s/v_s^0  \approx 1 - \left[6.55 - 3.52(1-B_2^*)\right]\phi + {\cal O}(\phi^2)
\end{equation}
where the normalized second virial coefficient~\cite{Russ89} is defined as $B_2^* \equiv B_2/B_2^{HS}$, with $B_2^{HS}$ the virial coefficient calculated with the effective HS radius of the colloids.  Eq.~(\ref{eq:eq1}) suggests that attractions should increase the sedimentation velocity, while added repulsions should decrease it compared to the pure HS case. Experiments on  dilute suspensions 
with inter-colloid attractions~\cite{Jans86,Plan09} or long-ranged electrostatic repulsions~\cite{Thie95} are consistent with this picture.
  However, these theories and experiments are only relevant for very dilute suspensions.  What happens to the average sedimentation velocity at larger volume fractions is not well understood, and almost nothing is known about the effect of attractions on velocity fluctuations at any particle density.

In this letter we address these questions by applying a mesoscopic simulation technique based on stochastic rotation dynamics (SRD)~\cite{Male99,Padd06}, that can successfully reproduce hydrodynamic fluctuations in steady-state sedimentation~\cite{Padd04} and has recently been shown to {\em quantitatively} describe colloidal sedimentation experiments, including  complex non-linear effects such as Rayleigh Taylor instabilities~\cite{Wyso09}.  The accuracy with which SRD was shown to reproduce colloidal experiments gives us confidence in the predictions of  our simulations with inter-particle attractions.  
We are able to go beyond the dilute regime and find that,  for short-ranged attractive potentials, Eq.~(\ref{eq:eq1}) is accurate up to about $\phi \approx 0.05$, but breaks down for higher packing fractions.   We measure, for the first time, the effect of attractions on velocity fluctuations  in the unscreened regime, and find that the range of the hydrodynamic swirls is greatly increased by attractions, with a maximum around $\phi \approx 0.07$ for stronger attractions. We link the increase in both average sedimentation velocity and in the size of the hydrodynamic swirls  to a complex interplay between the 
aggregation, fragmentation and sedimentation of transient clusters.

In our simulations, the colloid-colloid interaction  was modeled by a classic DLVO~\cite{Russ89} potential: $V_{cc}(r)=V_{HS}(r)+V_{vdW}(r)+V_{DH}$. The first term is a repulsive HS like contribution
of the WCA form:  $\beta V_{HS}(r)=10\left[ (\sigma /r)^{2n}-(\sigma /r)^{n}+1/4 \right]$ for $r \le 2^{1/n}\sigma$ and 0 for $r > 2^{1/n}\sigma$, where $n=24$ and $\sigma =\sigma_{cc}$, the colloidal HS radius.  The second and third terms are the short-range van der Waals attraction and the repulsive Debye-H\"uckel-like contribution respectively ~\cite{Russ89,Pell03}. In order to overcome the singularity of the van der Waals contribution at contact, we also introduced a cut-off distance (the so-called Stern layer) given by $\delta = 0.048\sigma_{cc}$~\cite{Pell03}. Keeping the Debye screening length fixed at $\kappa=8.96/\sigma_{cc}$, we varied the Hamaker constant and particle charge to obtain four different  potentials with an attractive minimum at short inter-particle distance. The normalized second virial coefficients were $B_2^* =-0.063$, $-0.507$, $-1.044$ and $-1.416$ respectively. Note that all these values are above $B_2^*=-1.5$ in order to avoid
fluid-fluid phase separation of the colloidal system~\cite{Vlie00}. 

Brownian fluctuations and HI are induced by including  SRD fluid particles in the simulation.  These particles interact with each other  through an efficient coarse-grained collision step that conserves mass, energy and  momentum, so that the Navier Stokes equations are recovered at the macroscopic level~\cite{Male99}; note that  in the literature this method is also called multiple particle collision dynamics and has been widely applied to soft-matter simulations~\cite{Gomp09}.   The colloids are coupled to the SRD fluid through an interaction of the WCA form with $\beta V_{cf}(r)=10\left[ (\sigma /r)^{2n}-(\sigma /r)^{n}+1/4 \right]$ for $r \le 2^{1/n}\sigma$ and 0 for $r > 2^{1/n}\sigma$, where $n=6$ and $\sigma =\sigma_{cf} =  0.465\sigma_{cc}$, and the ensuing equations of motion are updated with a standard Molecular Dynamics algorithm for the colloid-colloid and the colloid-fluid interactions, and with a coarse-grained SRD collision step for the fluid-fluid interactions
 .  The colloid-fluid diameter $\sigma_{cf}$ is lightly smaller than $0.5 \sigma_{cc}$ to avoid spurious depletion forces between the colloids~\cite{Padd04,Padd06}.  This mesoscopic simulation technique has been shown to reproduce the correct low Reynolds (Re) hydrodynamic flow behaviour with an effective hydrodynamic radius of $a \approx 0.8 \sigma_{cf}$, as well as the correct thermal Brownian fluctuations and diffusion for colloidal suspensions.  We refer the reader to Refs.~\cite{Padd04,Padd06} for further technical details and a justification of  our SRD  parameter choice.

The simulations were performed by placing $N_c=8-819$ colloids in a box of sizes $L_x=L_y=16\sigma_{cf}$ and $L_z=48\sigma_{cf}$ with periodic boundary conditions. The number of solvent particles was $N=40V_{free}/\sigma_{cf}^3 \sim 4-5 \times 10^5$, where $V_{free}$ is the free volume left by the colloids. A gravitational external field $g$ is applied to the colloids in the $z$ direction in order to induce sedimentation.  After an initial transient time,  the system reaches   steady-state conditions, where the average sedimentation velocity $v_s$ is constant, the one-body particle spatial distribution is homogeneous, and no drift is observed.  To keep the centre of mass fixed, the downward volume flux of colloids {\bf is} compensated by an equivalent upward volume flux of fluid.   The simulation box sizes are small enough that we are still in the unscreened regime~\cite{Padd04}; larger simulations and possibly different  boundary conditions~\cite{Ladd02,Much04}  are necessar
 y in order to observe screening.  
    The particle Re number $\text{Re}= v_s a/\nu $, where $\nu$ is the kinematic viscosity, was kept at $\text{Re} \approx 0.08$, which is small enough for the system to be in the correct low Re number Stokesian regime.  Similarly, the P\'eclet number $\text{Pe} = v_s a /D_c$, with $D_c$ the colloid diffusion coefficient, was kept at $\text{Pe}=2.5$, so that thermal Brownian noise is non-negligible.

\begin{figure}
\center\resizebox{0.45\textwidth}{!}{\includegraphics{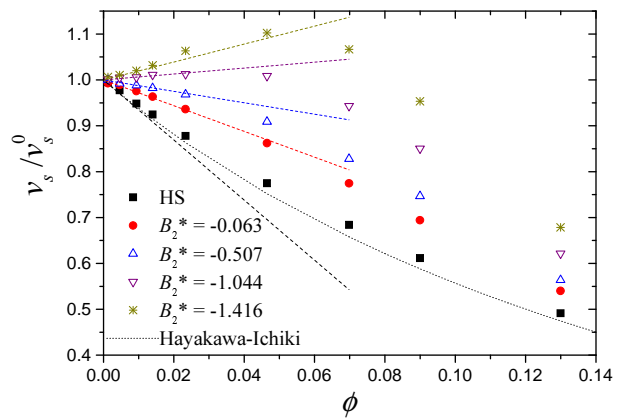}}
\caption{\label{Fig1} Average sedimentation velocity, $v_s$ as a function of the volume fraction $\phi = \frac43 \rho a^3$ for four different inter-particle attractions and for hard spheres. The dashed lines are Batchelor's predictions~(\ref{eq:eq1}) for the dilute limit. The dotted line is a prediction known to be accurate for hard spheres~\protect~\cite{Haya95}.}
\end{figure}

We begin our study with the effect of the inter-particle attractions on the average sedimentation velocity.  Fig.~\ref{Fig1} shows $v_s$ for various attractions and for different  packing fractions $\phi=\frac{4}{3}\pi \rho a^3$, where $\rho$ is the colloid number density.  Note that for these hydrodynamic effects, the correct radius to use is the hydrodynamic one.  
In the dilute limit, the simulation results are well described by  Batchelor's  prediction~(\ref{eq:eq1})
 with no free parameters.   For $B_2^*>-0.86$ the slope at low-$\phi$ is negative, but for $B_2^*<-0.86$ the attractions become strong enough to overcome the backflow-induced velocity reduction and give rise to a positive slope. These results suggest that, at least for short ranged  attractions, the exact  potential details are unimportant and $v_s/v_s^0$  is controlled by $B_2$.

At larger packing fractions the Batchelor prediction breaks down.  Interestingly, for more attractive systems 
 ($B_2^*<-0.86$) we observe a clear maximum in $v_s$ v.s. $\phi$ that becomes more pronounced for stronger attractions, as predicted by theory~\cite{Gill04}. To the best of our knowledge, this is the first time that a non-monotonous dependence on $v_s$ has been observed in simulations.

\begin{figure}
\center\resizebox{0.45\textwidth}{!}{\includegraphics{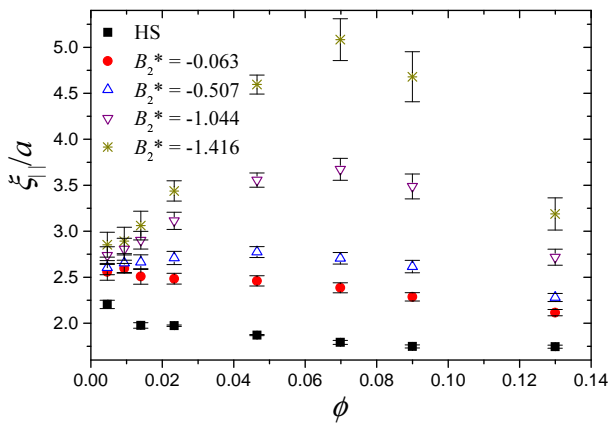}}
\caption{\label{Fig2} Correlation length $\xi_{||}$  parallel to the sedimentation direction for velocity swirls,  as a function of packing fraction $\phi$ for  different inter-particle attractions and for pure HS.  Attractions strongly enhance the size of the velocity swirls.}
\end{figure}

Having established that attractions significantly influence the sedimentation velocity, we now turn to the velocity fluctuations of the particles around the average, $\delta v=v-v_s$. 
Here we focus on spatial correlations in the $z$-direction, defined as $
C_z(\vec{r})=\left< \delta v_z(0)\delta v_z(\vec{r})\right> / \left< \delta v_z(0)^2 \right>,$
where $\left< ... \right>$ is the average over time and over all colloids. The distance vector $\vec{r}$ may be taken parallel to the sedimentation, $C_z(z)$, or perpendicular to it, $C_z(x)$. The correlation function $C_z(z)$ exhibits an exponential decay $C_z(z)=\text{exp}(-z/\xi_{||})$, where $\xi_{||}$ is the correlation length in the direction parallel to the sedimentation. It provides a  measure of the size of the hydrodynamic swirls. $C_z(x)$ decays as well,  but also shows  an anti-correlation region where the swirl moves in the opposite direction.  The qualitative shape and decay of the swirls is similar to that observed in experimental observations~\cite{Segr97} and simulations~\cite{Padd04} for HS particles.  However,  as can be seen in Fig.~\ref{Fig2}, the correlation length  $\xi_{||}$ is  greatly enhanced by 
the attractions.  Whereas for weaker attractions $\xi_{||}$ decreases with $\phi$, similar to what is observed for HS particles~\cite{Segr97,Padd04}, for more strongly attractive systems the correlation length first increases with $\phi$ and then decreases, giving rise to a maximum around $\phi \approx 0.07$ for $B_2^* < -1$. A similar non-monotonic behavior with $\phi$  is observed for $C_z(x)$ (not shown), i.e. the depth of the anticorrelation region grows for increasing particle attractions and, for the particular case of $B_2^*<-1$, it shows a maximum at intermediate colloidal density. 

\begin{figure}
\center\resizebox{0.45\textwidth}{!}{\includegraphics{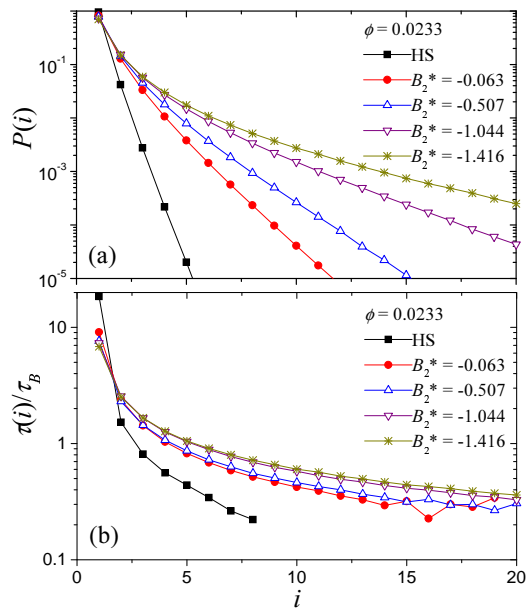}}
\caption{\label{Fig3} 
(a) The probability of finding a transient cluster of size $i$ and (b) the average cluster life-time (normalized by the Brownian time $\tau_B=a^2/D_{col}$) as a function of $i$ for hard spheres and for four different inter-particle attractions. The particle volume fraction is $\phi=0.0233$.}
\end{figure}

Although the attractions in our system are not strong enough to form permanent clusters, they do enhance the probability for particles to cluster together in a transient fashion.  This effect can be clearly seen Fig.~\ref{Fig3} where we plot the probability distribution of transient clusters $P(i)$ and their average life-time $\tau(i)$ as a function of the cluster size, $i$.  
In order to distinguish whether a pair of particles belong to the same cluster 
 or not, a cut-off distance $r_{cut}=1.06\sigma_{cc}$ was used, which is a reasonable estimate of the range of the attractive potential well.  We checked that our results do not qualitatively depend on the exact cutoff distance $r_{cut}$.  As the attraction strength increases, both the probability of finding clusters ($P(i>1)$) and the average cluster life-time $\tau(i)$ increases.   The gravitational force on a cluster increases linearly with the number of particles inside the cluster, but the friction increases roughly linearly with the radius of gyration of the cluster.  Since the latter typically increases less quickly than the former, larger clusters sediment faster than smaller ones, an effect we observe by tracking the clusters in time. Furthermore, we find that stronger interactions lead to slightly more compact clusters, with a smaller radius of gyration and so even larger sedimentation velocities. This cluster picture compliments the observation~\cite{Russ89} that a
 ttractions decrease the contribution of the hydrodynamic backflow term used  to derive Eq.~(\ref{eq:eq1}).
 Together, these effects help explain why at a fixed  $\phi$, increasing the strength of the attractions increases the average sedimentation velocity $v_s$. 
 Furthermore, faster clusters with a longer life-time are able to move along larger distances with a roughly constant velocity. This leads to the propagation of the correlations along larger distances and so, to an increase of $\xi_{||}$ which tracks the increase in the size of the swirls.

\begin{figure}
\center\resizebox{0.45\textwidth}{!}{\includegraphics{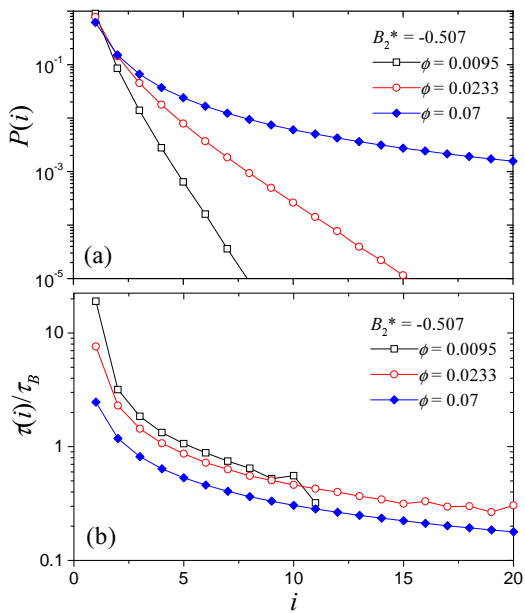}}
\caption{\label{Fig4} The plots show (a) the probability of finding a transient cluster of size $i$ and (b) the average cluster life-time (normalized by the Brownian time $\tau_B$) as a function of $i$ for three different particle volume fractions, $\phi$. The normalized second virial coefficient is $B_2^*=-0.507$.}
\end{figure}

 In Fig.~\ref{Fig4} we show how the cluster distributions and life-times change with $\phi$ for a fixed attraction.  Although we only show one value of $B_2$, the curves show the same qualitative behaviour for other values of the attraction we studied.  For larger $\phi$, particles are on average closer to each other, and so transient clusters are more likely to occur, increasing $v_s$.   On the other hand, 
the clusters have a shorter life-time for larger $\phi$, presumably because the collision rate with other particles increases.   This effect will decrease $v_s$.
The backflow also increases with $\phi$, decreasing the average $v_s$.  At low $\phi$ and stronger attractions, the enhanced transient cluster formation wins out, but at higher $\phi$ the shorter cluster life-times as well as the enhanced backflow reduce $v_s$ again.
 In summary then,    
 the competition between larger clusters and shorter life-times may explain the more complex dependence of the sedimentation velocity  and velocity fluctuations on volume fraction that we observe for attractive particles.  
  
Finally, this study also raises a number of further questions.  Firstly,  our simulations have been for a low Pe number, but it would be interesting to see what happens for larger Pe numbers.   Preliminary simulations up to Pe$=15$ show that, in contrast to the pure HS case where the $v_s/v_s^0$ v.s.\ $\phi$ curves all scale onto the same form for Pe numbers in this range~\cite{Padd04}, we observe that   $v_s/v_0$ is  slightly reduced at  higher  Pe number.  A cluster analysis reveals that larger clusters are slightly less likely to occur at higher Pe numbers, most likely because the enhanced shear forces break them up, leading to a lower sedimentation velocity. What happens in the non-Brownian infinite Pe number limit remains to be investigated. 

Secondly, it would be very interesting to study larger box sizes, to see whether attractions change the crossover from the screened to the unscreened regime. Thirdly, for even stronger attractions ($B_2^* < -1.5$) permanent clusters should begin to form. These will then sediment more quickly than monomers. But as they grow and accelerate, at some point the shear forces should break them up again.  Such a rich interplay between aggregation and hydrodynamics should lead to new steady states with a cluster population that depends on the attraction strength. New simulations are planned to address these questions.

The authors thank the Spanish Ministerio de Educaci\'on y Ciencia (project MAT2006-12918-C05-01), the Junta de Andaluc\'{\i}a (project P07-FQM02517), the Royal Society (London) and the Netherlands Organization for Scientific Research (NWO) for financial support.



\end{document}